\documentclass[12pt]{article}
\usepackage[T1]{fontenc}
\usepackage{graphicx}
\usepackage{geometry}
\usepackage{makecell}  
\usepackage{booktabs}
 \usepackage{url}

\title{Evaluating Personalized Beneficial Interventions in the Daily Lives of Older Adults Using a Camera}

\author{
Longfei Chen, Christopher Lochhead, Robert B. Fisher,\\
Nusa Faric, Jacques Fleuriot, Subramanian Ramamoorthy \\
\\
\textit{School of Informatics and Advanced Care Research Centre,} \\
\textit{The University of Edinburgh,} \\
\textit{10 Crichton St, Edinburgh, EH8 9AB, United Kingdom}
}

\date{}

\begin{document}

\maketitle

\begin{abstract}
Beneficial daily activity interventions have been shown to improve both the physical and mental health of older adults. However, there is a lack of robust objective metrics and personalized strategies to measure their impact. In this study, two older adults aged over 65, living in Edinburgh, UK, selected their preferred daily interventions (mindful meals and art crafts), which are then assessed for effectiveness. The total monitoring period across both participants was 8 weeks. Their physical behaviours were continuously monitored using a non-contact, privacy-preserving camera-based system. Postural and mobility statistics were extracted using computer vision algorithms and compared across periods with and without the interventions. The results demonstrate significant behavioural changes for both participants, highlighting the effectiveness of both these activities and the monitoring system.
\end{abstract}

\vspace{1em}
\noindent\textbf{Keywords:} Older adults, daily activity interventions, objective monitoring, privacy-preserving

\section{Introduction}
Studies have shown that older adults spend approximately two-thirds of their waking hours, 8 to 12 hours per day, engaged in sedentary activities~\cite{ref15}, which significantly increases the risk of various health conditions~\cite{ref14}. Promoting physical activity and healthy behaviours is crucial to enhancing the quality of life for older adults~\cite{ref5}; it can reduce blood glucose levels, lower blood pressure, and help prevent disease~\cite{ref7}.

Numerous studies have explored daily beneficial interventions to promote older adults’ activeness and quality of life. For example, physical activity programs~\cite{ref1,ref2}, such as structured walking programs and supervized exercises, have been shown to enhance physical functioning and reduce the risk of mobility disabilities~\cite{ref6}. Motivational prompts such as music playlists~\cite{ref18}, artwork~\cite{ref8}, social interaction~\cite{ref13}, and nature-based activities~\cite{ref11} improve the mental health of older adults~\cite{ref10,ref12}. Mindful eating~\cite{ref16} has also been shown to help with weight management and reduce depression, perceived stress, and more~\cite{ref17}.

Most of these studies use standardized measures of quality of life or well-being, including mood and life satisfaction, with both subjective well-being components and clinical depression and anxiety scales~\cite{ref3}. However, a significant challenge lies in the lack of uniform and objective metrics to effectively measure and compare the outcomes of various intervention strategies~\cite{ref2}. 
{Additionally, most of the interventions are designated to a group of participants rather than personalized to individual.}

Wearable sensors are commonly employed to objectively measure older adults’ behaviours, with devices placed on the wrist~\cite{ref9}, waist~\cite{ref24}, or legs~\cite{ref25} to capture data for various of metrics, such as step counts. While these devices offer valuable data, they may not be suitable for everyone and can carry stigma~\cite{ref19}. Some older adults, particularly those with cognitive impairments, may find it challenging to remember to wear and charge these devices consistently~\cite{ref20}.

\begin{figure}[t]
\centering
\includegraphics[width=.9\textwidth]{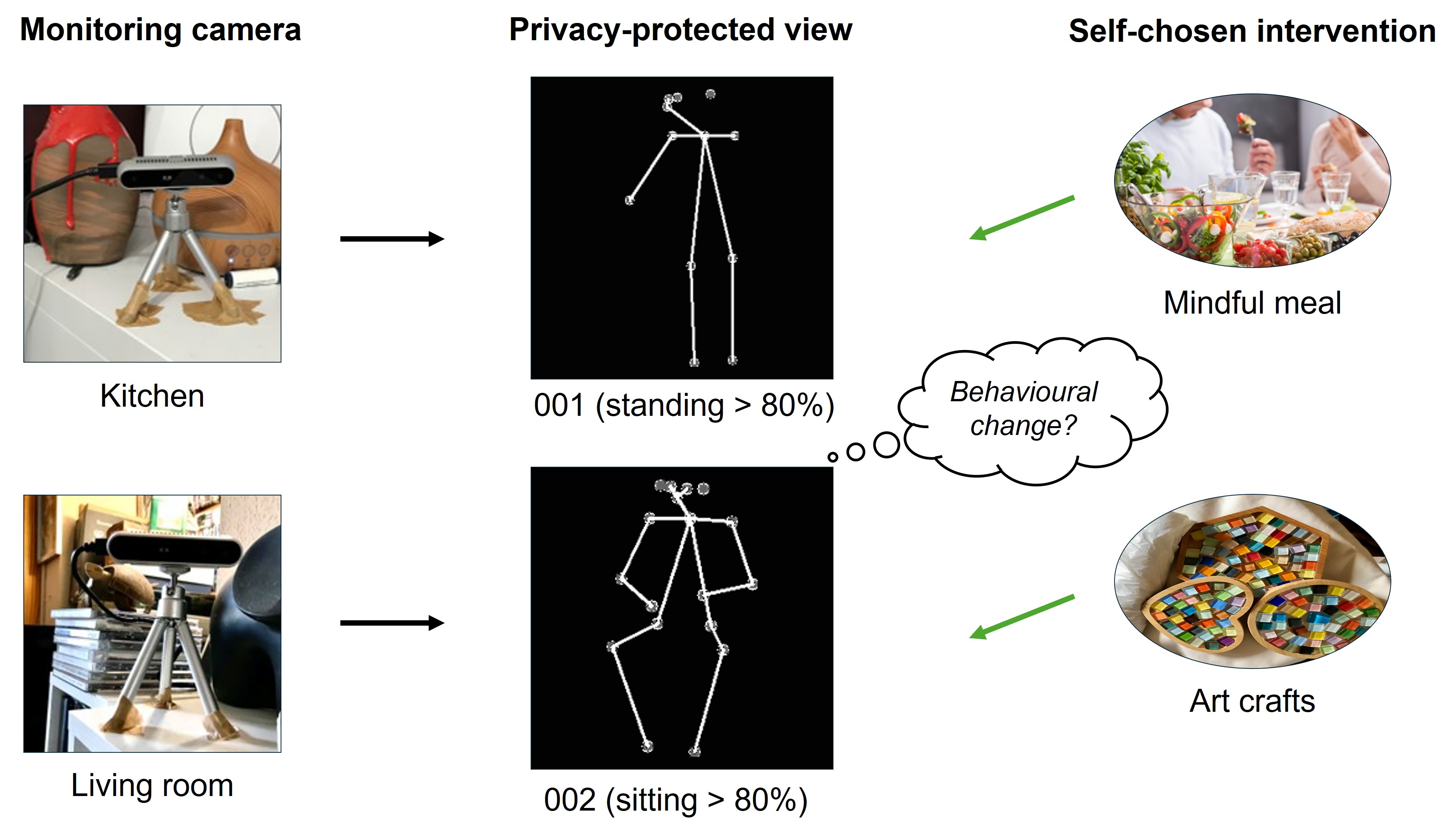}
\caption{
Monitoring systems were deployed in two home environments to capture body motion data. An RGB-D camera (Intel RealSense D415) and a compact computer (Jetson Orin Nano) were positioned approximately 2 metres away to monitor key areas of each room. Visual input was processed in real time to generate 2D pose representations, capturing human presence, movement, and posture, without saving any images or videos in order to preserve privacy. The aim was to observe behavioural changes in response to the selected intervention activities.
Two older adult participants (001 and 002) selected mindful meals and art crafts as their preferred intervention activities and were monitored in the kitchen and living room, respectively. Participant 001 was observed to spend over 80\% of their time standing in the kitchen, whereas Participant 002 spent over 80\% of their time sitting in the living room.} \label{pipe}
\end{figure}

As an alternative, camera-based monitoring systems provide accurate assessments of physical activity without the need for individuals to wear any devices. One system can monitor whole-body movements and has been accepted at clinical levels~\cite{ref21,ref22}. However, considerations regarding privacy and acceptance among older adults remain critical for their implementation~\cite{ref23}.

This study investigates how personalized interventions can encourage older adults to remain active and improve their health. Specifically, we implemented two interventions selected by the participants themselves: mindful meals and art crafts. Our technology provides an automatic and privacy-preserving method for monitoring domestic behaviour, using a camera-based system to track body motion and behavioural changes during these activities (see Fig.~\ref{pipe}). The analysis focuses on evaluating the impact of these interventions by examining changes in participants' behaviours and motion levels, offering a comprehensive understanding of their effects on daily life. The results show that both participants exhibited significant behavioural changes during the intervention periods, including shifts in daily routines and measurable differences in physical activity patterns.

The main contributions of this paper are:
\begin{enumerate}
    \item A novel integration of a privacy-preserving, camera-based monitoring system to continuously observe self-chosen personalized intervention events in real-world settings, including mindful meals and art crafts;
    \item The effectiveness of these interventions on behavioural changes in older adults is assessed through objective data and camera-derived postural and mobility metrics.
\end{enumerate}

\section{Methods}
The purpose of this study is to evaluate the effectiveness of personalized beneficial activities in the daily lives of older adults. First, based on a Patient and Public Involvement and Engagement program, the interviewed older adults identified a list of six potentially beneficial intervention events that they consider beneficial for themselves, including \textit{music playlists}, \textit{artwork}, \textit{getting into nature}, \textit{in-person visits}, \textit{mindful meals}, and \textit{drinking reminders}. Each participant then selected their preferred intervention activity from this list, and two monitoring periods were conducted. \footnote{The study was approved by the School of Informatics Research Ethics Committee, and individual informed consent was obtained.} 

During the first period, participants followed their usual routines without any intervention (the baseline phase). In the second period, participants actively engaged in their chosen intervention activity on a daily basis {(see Section~\ref{context} for further details).} Each monitoring phase lasted between 1 to 3 weeks, depending on the participant’s preference. 

To facilitate data collection, a compact motion measurement system was deployed (see Fig.~\ref{pipe}). This system includes an RGB-D camera, a small computer processor, and a power socket. It is designed to monitor participants' full body movements within a designated room. The camera is positioned to cover key activity areas. The monitoring is fully automated, {unobtrusive}, and requires no interaction. To preserve privacy, all video data is processed in real time, and only anonymized features are saved.

Specifically, the following features are captured: {(a) human appearance}, {(b) body movement characteristics}, and {(c) body postures}.\\
\textbf{(a) Human appearance:} Human presence is detected using a lightweight 2D pose estimation method~\cite{ref26} and YOLOv5 object detection~\cite{ref27}. If both detectors indicate over 50\% probability of a person being present, the timestamp is recorded as a human appearance.\\
\textbf{(b) Body movement characteristics:} Three metrics are calculated to describe body movement: \textit{Inactivity}, \textit{Movement Scale}, and \textit{Movement Speed}. Significant motion (i.e., active pixels) is detected via pixel-level changes across all RGB channels~\cite{ref28}. 2D dense optical flow is used to estimate body motion speed, achieving a precision of up to 2.5 pixels RMSE per frame~\cite{ref29}. \textit{Inactivity} is defined as the presence of a person with no detectable significant motion in the body area. \textit{Movement Scale} is defined as the ratio of the area covered by moving parts of the body (bounding box of active pixels) to the total body area (as detected by YOLOv5). \textit{Movement Speed} refers to the average 2D motion speed per frame computed using optical flow. Detailed definitions of these metrics are available in~\cite{ref30}.\\
\textbf{(c) Body postures:} Sitting and standing time are key indicators of healthy lifestyle habits. We employ a pose-based classification method where 2D posture images are classified into three categories: “sitting”, “standing”, and “others.” A ResNet18 model~\cite{ref31} is trained on manually labelled data (10,000 binary 2D posture images, augmented to 74,408 images) from seven older adults captured at their homes, and tested on 36,640 images from the same population. The classifier achieved an accuracy of 0.998 for sitting and 0.997 for standing in this population.

{Finally, features extracted at the frame level are aggregated into hourly averages and then summarised for each monitoring day. To evaluate the effect of the interventions, a \textbf{paired t-test} is used to compare key behavioural features between the normal and intervention periods for each participant. This test is appropriate when measurements are taken from the same individual under two conditions (i.e., before and during the intervention), and the monitoring periods are of equal length~\cite{ref32}, as in this study. In particular, we pair the data by corresponding days in the normal and intervention phases, as well as match weekdays (e.g., Mon. to Mon., Tue. to Tue.), to assess whether the mean differences in characteristics are statistically significant. These characteristics include daily appearance duration, movement intensity, posture ratios, and hourly appearance data across all days. We acknowledge, however, that the relatively small sample sizes (8 paired days for Participant 001 and 21 paired days for Participant 002) may introduce bias and limit the statistical power of the test.}


\section{Results}
\subsection{Monitoring Report}

Participant 001 is a 67-year-old male (BMI 31.1) living alone. He selected the mindful meal intervention, which involved intentionally focusing on food choices, preparation, and eating habits to encourage healthier dietary behaviours. He was monitored over a total of 16 days, i.e., 8 days without intervention as a baseline (referred to as the ‘normal’ period), followed by 8 days during the intervention. The monitoring camera was installed in a corner of the kitchen above the fridge, covering the entire kitchen area to unobtrusively capture his daily routines and behaviour changes.

Posture detection results show that the participant spent most of their time standing while in the kitchen, both with and without the intervention (normal: mean 85.5\%, SD 13.4\%; intervention: mean 79.3\%, SD 19.7\%). There was a slight decrease in standing time (increase in sitting time) during the mindful meal intervention. No significant differences were observed in the duration of standing or sitting in the kitchen between normal and intervention periods (p = 0.344 and p = 0.593, respectively).

\begin{figure}
\centering
\includegraphics[width=.9\textwidth]{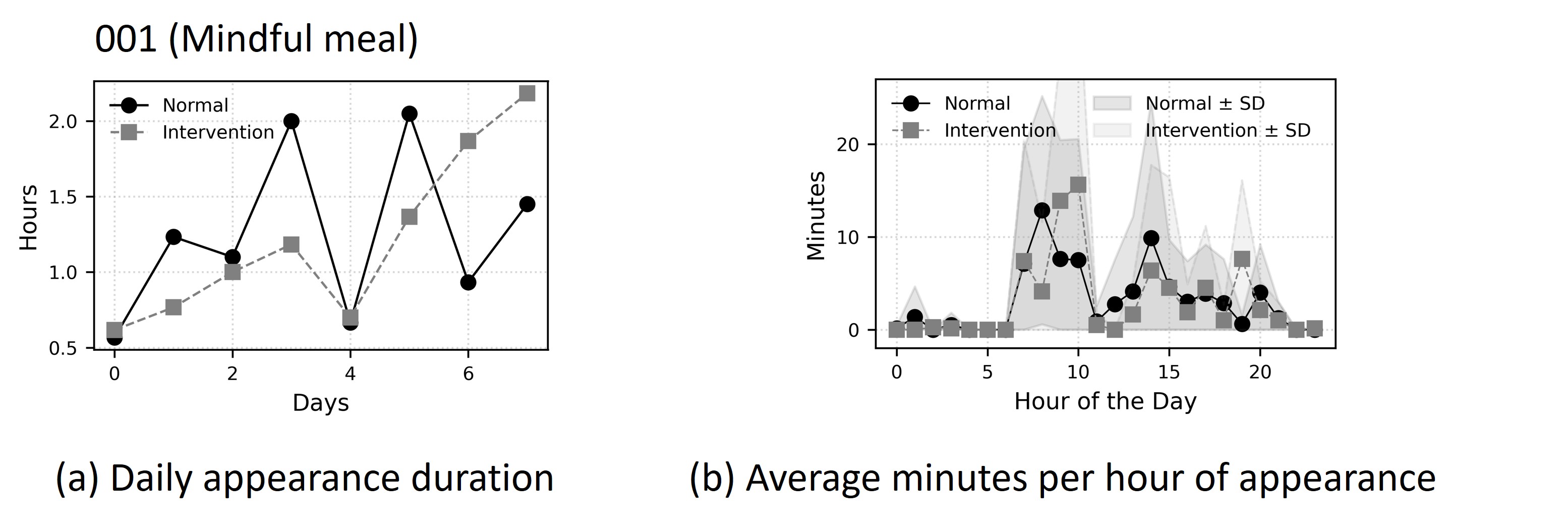}
\caption{Comparison of kitchen usage patterns between normal and intervention (mindful meal) periods for Participant 001. (a) Total duration of kitchen appearances per day over 8 monitored days. (b) Average duration of kitchen presence per hour for 24-hour across all days. The intervention (mindful meal) period shows changes in temporal kitchen usage, including slightly reduced nighttime visits and a shift in peak morning activity.
} \label{time1}
\end{figure}

As shown in Fig.~\ref{time1}(a), during the normal period, the total time spent in the kitchen each day ranged from 0.5 to 2 hours. During the intervention period, this daily total gradually increased. However, the overall difference between the normal and intervention periods was not statistically significant.
By analysing the average time spent in the kitchen each hour (00:00–-23:00), shown in Fig.~\ref{time1}(b), we observed that on normal days, the participant occasionally visited the kitchen during the night (00:00–-05:00). These late-night visits decreased by 81.3\% during the intervention period compared to normal days. 

A shift in breakfast time was observed, from 7:00–-9:00 (with a peak at 8:00 and an average duration of around 13 minutes) during normal days to 9:00-–11:00 (peak at 10:00, averaging around 16 minutes) during the intervention period. In the afternoons, the participant spent an average peak of 10 minutes in the kitchen at 14:00 on normal days, while the duration at that time slightly decreased during the intervention. Instead, a new peak emerged at 19:00 during the intervention period, with an average duration of 8 minutes, suggesting a potential shift in eating habits toward evening meals. Overall, the distribution of eating times during the intervention appeared more concentrated.

\begin{figure}
\centering
\includegraphics[width=.95\textwidth]{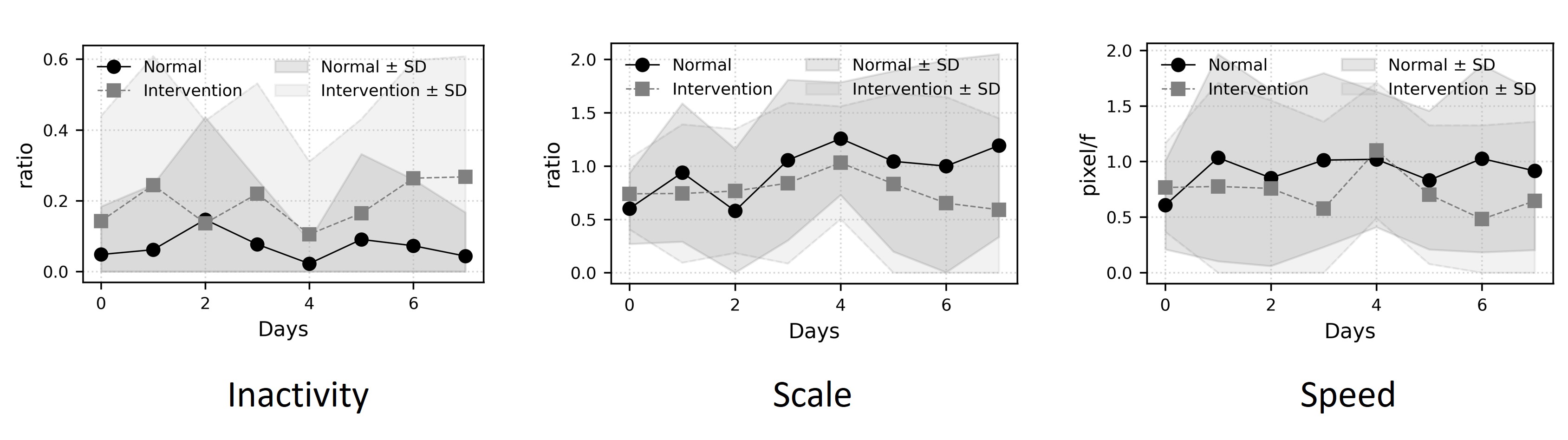}
\caption{Daily changes in body movement features 
for Participant 001 under normal and mindful meal intervention conditions. Left: Ratio of inactivity, showing a significant increase during the intervention period (p = 0.003). Middle:  Movement scale, indicating a trend toward reduced body movement amplitude during intervention (p = 0.077). Right: Movement speed, which decreased significantly during intervention days (p = 0.049).
} \label{motion1}
\end{figure}

Despite these trends, no statistically significant differences were found between the two conditions.

In terms of body movement features, as shown in Fig.~\ref{motion1}, the participant was significantly more inactive during the intervention period (p = 0.003). Movement speed decreased significantly (p = 0.049). The overall movement scale was smaller (p = 0.077), {though not statistically significant}. These findings indicate a notable shift in eating behaviour during the intervention period, suggesting the mindful meal approach influenced how the participant moved and behaved in the kitchen.

\begin{figure}[t]
\centering
\includegraphics[width=.9\textwidth]{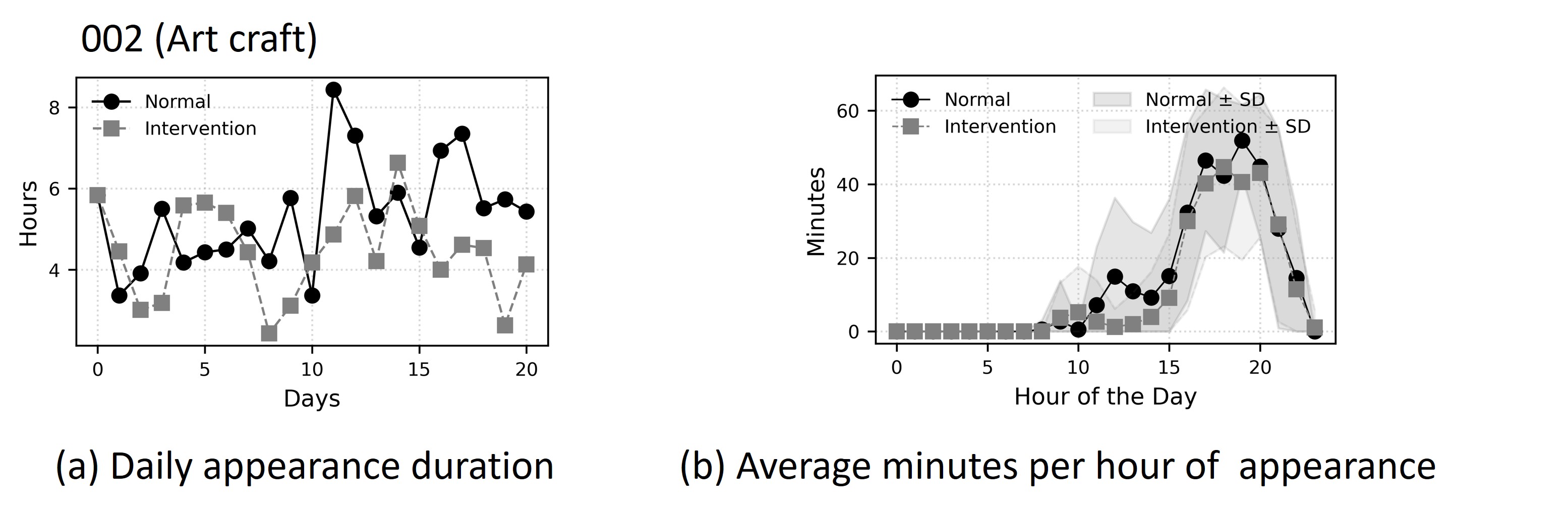}
\caption{Comparison of living room usage patterns between normal and intervention (art crafts) periods for Participant 002. (a) Total daily duration of living room appearances over 21 monitored days. (b) Average duration of living room presence per hour for 24-hour across all days.
} \label{time2}
\end{figure}

Participant 002 is a 75-year-old female (BMI 36.1) living alone. She was monitored for 21 days under normal conditions and another 21 days during an art and crafts intervention. During the intervention, she was provided with her preferred crafting activities. 
A camera was positioned on a shelf near the television, facing her favorite couch, with the primary focus on monitoring her seated behaviour in the living room. She actively participated in the intervention, incorporating the art crafts into her daily routine during the monitored period.

Posture detection results show that the participant spent most of her time sitting (normal: mean 83.6\%, SD 11.4\%; intervention: mean 90.1\%, SD 5.7\%), predominantly from the afternoon to evening.

\begin{figure}[htbp]
\centering
\includegraphics[width=.95\textwidth]{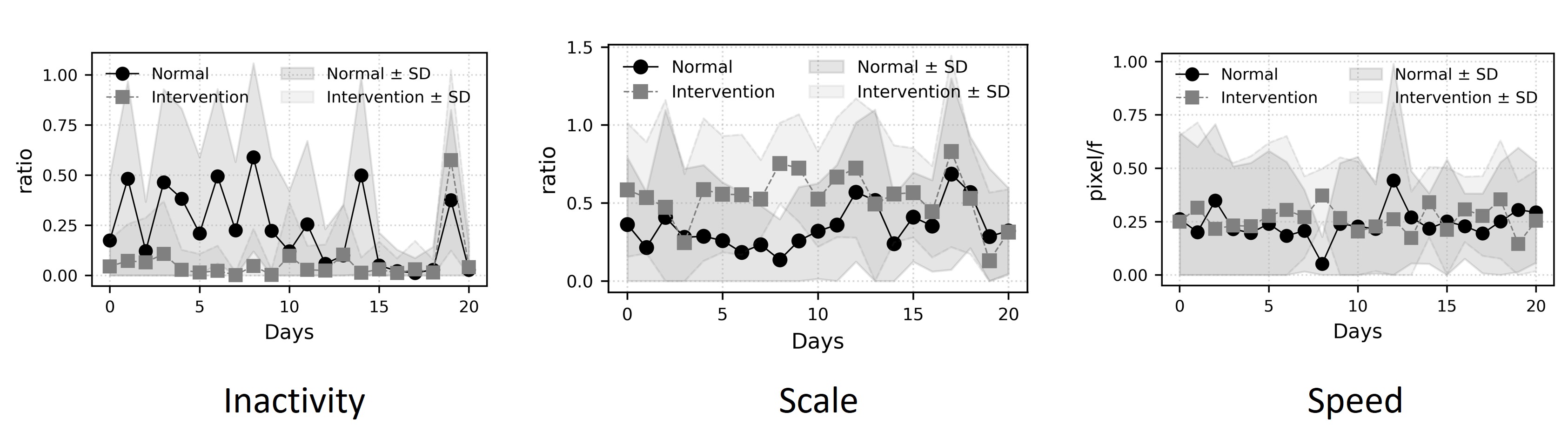}
\caption{Comparison of body movement features in the living room between normal and intervention (art crafts) periods for Participant 002. Left: Inactivity ratio per day (p = 0.001). Middle: Body movement scale, reflecting motion amplitude (p = 0.0001). Right: Body movement speed (p = 0.399). The results indicate significantly reduced inactivity and increased movement scale during the intervention, with no significant change in movement speed.
} \label{motion2}
\end{figure}

As shown in Fig.~\ref{time2}(a), the duration of living room appearances exhibits high day-to-day variability. The average appearance time in the living room during the daytime decreased by approximately 1 hour during the intervention (normal: mean 5.36 h, SD 1.30 h; intervention: mean 4.47 h, SD 1.11 h). As shown in Fig.~\ref{time2}(b), usage was primarily concentrated between 08:00 and 20:00, with two minor peaks around midday and early evening. The most notable reductions occurred between 12:00–-13:00 and at 19:00, while differences at other hours were relatively minor.

Both the total daily appearance duration (p = 0.019) and the average appearance per hour across the day (p = 0.021) differed significantly between the normal and intervention phases.

As shown in Fig.~\ref{motion2}, following the intervention, inactivity significantly decreased (p = 0.001), suggesting more engaged or purposeful sitting behaviour. Movement scale (indicative of body motion amplitude) increased significantly (p = 0.0001), which may reflect more dynamic upper-body movements during the art activity. Movement speed showed no significant change (p = 0.399), indicating that while movements became larger, they did not become faster.

\begin{table}[ht]
\centering
\footnotesize
\caption{{Comparing p-values of differences in statistics of behavioural features between normal and intervention periods. Bold values indicate statistically significant differences ($p < 0.05$).}}
\resizebox{\textwidth}{!}{%
\begin{tabular}{lccccccc}
\toprule
\makecell[l]{\textbf{\textit{T}-test ($p$-value)}} & 
\makecell{\textbf{Standing}  } & 
\makecell{\textbf{Sitting}  } & 
\textbf{Inactivity} & 
\makecell{\textbf{Scale}} & 
\makecell{\textbf{Speed}} & 
\makecell{\textbf{Appearance} \\ \textbf{duration}} & 
\makecell{\textbf{Appearance} \\ \textbf{min. per h.}} \\
\midrule
\makecell[l]{P1 (16 days, DoF = 7)} & 
0.344 & 0.593 & \textbf{0.003} & 0.077 & \textbf{0.049} & 0.863 & 0.888 \\
\makecell[l]{P2 (42 days, DoF = 20)} & 
0.517 & \textbf{0.020} & \textbf{0.001} & \textbf{0.0001} & 0.399  & \textbf{0.019} & \textbf{0.021} \\
\bottomrule
\end{tabular}
}
\label{tab:paired_ttest}
\end{table}

{The comparison of behavioural characteristics between the normal and intervention periods for both participants is summarised in Table~\ref{tab:paired_ttest}.}

\subsection{Subjective Feedback and Contextual Factors} \label{context}
The interventions were self-driven, allowing participants to engage in their chosen activity based on their routines. No fixed schedule or automated reminders were used. Instead, participants were encouraged to carry out the activity as frequently as they felt comfortable during the intervention period, and to log the dates and times of each session. This design choice was made to minimise disruption to natural behaviour.
Additionally, in our analysis, we treat all days within both baseline and intervention as complete units, focusing on the evaluation of the overall behavioural impact of the intervention rather than the momentary responses to individual activity sessions.

Participant 001 is a highly organized male who enjoys solitary hobbies such as assembly, DIY, and reading. He selected the mindful meal intervention, which involved paying close attention to the types of food purchased and consumed, with a focus on slower eating and an emphasis on low-sugar, low-carbohydrate options to support weight management and blood sugar control. According to his personal records, his daily meal schedule was highly variable, with 3 to 4 meals consumed at different times throughout the day. He reported being satisfied with the mindful meal intervention, paying special attention to his food choices and consistently recording mealtimes. This heightened awareness may account for the observed increase in eating duration and a reduction in body movement during the intervention period, likely reflecting a more “mindful” approach to meals. Overall, he reported feeling emotionally stable throughout both the normal and intervention periods.

Participant 002 is a hospitable woman living alone, who primarily uses the downstairs living room for daily activities. She chose the art crafts intervention due to a strong personal interest in creative activities such as sewing and jewellery-making at community centres. She specifically selected diamond and mosaic crafts, both of which were new experiences for her. Given a knee condition that limits her ability to walk daily, crafting provided an engaging and accessible alternative. She consistently scheduled her craft sessions between 17:00 and 19:00. Notably, the intervention did not lead to an increase in sitting time, but a clear rise in active behaviour was observed. She described the experience as enjoyable and relaxing, helping her feel calm and content. Throughout the intervention period, she consistently reported feeling happy.

\section{Conclusion}
Both participants exhibited statistically significant behavioural changes following the implementation of personalized intervention activities. These changes were evident in their patterns of presence, postural behaviours, and movement characteristics. Specifically, Participant 001 demonstrated altered kitchen routines and a reduction in movement intensity during the mindful meal intervention. In contrast, Participant 002 showed a marked decrease in inactivity and an increase in movement magnitude during the art crafts intervention. 
{These findings suggest that personalized daily activities might effectively foster behavioural change engagement among older adults (i.e., as we unobtrusively monitored in real-world settings).}

The ability to observe these behavioural changes through {extended,} automatic, non-invasive, and privacy-preserving monitoring 
{is a potentially valuable tool} for healthcare professionals as well as for the individuals themselves. Users can independently track changes in their own behaviour without the need for ongoing management of the monitoring system, providing a practical and sustainable approach to long-term health monitoring. 
The main limitations of this pilot study lie in the small number of participants and the relatively short monitoring period. Additionally, personalized intervention strategies, while potentially more effective for individuals, are inherently more challenging to generalize across broader populations. Future work will involve expanding the study population to evaluate the wider applicability and effectiveness of personalized interventions across diverse groups.

%

\section*{Acknowledgements}
This research was funded by the Legal \& General Group (research grant to establish the independent Advanced Care Research Centre at the University of Edinburgh).  
The funder had no role in the conduct of the study, interpretation, or the decision to submit for publication.  
The views expressed are those of the authors and not necessarily those of Legal \& General.
This research was also funded by the Moray Endowment Fund, The University of Edinburgh.

\section*{Conflict of Interest}
The authors declare that they have no competing interests relevant to the content of this article.

%
%
%
%

\end{document}